\documentclass[11pt]{article}
\usepackage{moriond,epsfig}

\def\be{\begin{equation}}
\def\ee{\end{equation}}
\def\bea{\begin{eqnarray}}
\def\eea{\end{eqnarray}}

\begin{document}
\vspace*{4cm}
\title{NEW RESULTS ON DIRECT CP VIOLATION IN CHARGED KAON DECAYS BY NA48/2}

\author{ G. LAMANNA\footnote{On behalf of NA48/2 Collaboration: Cambridge, CERN, Chicago, Dubna, Edinburgh, Ferrara, Firenze, Mainz, Northwestern,
 Perugia, Pisa, Saclay, Siegen, Torino, Vienna.} }

\address{Department of Physics \& INFN Pisa, 3 Largo Pontecorvo,\\
56127 Pisa, Italy}

\maketitle\abstracts{The NA48/2 result, based on the data collected during the 2003 run, on direct CP violation 
in $K^{\pm}\rightarrow\pi^{\pm}\pi^0\pi^0$ decay is presented. The main goal of the experiment is to reach a sensitivity at level of $10^{-4}$ 
in the measurement of the charge asymmetry parameter $A_g=(g^+-g^-)/(g^++g^-)$, where $g$ is the \emph{linear slope} of the Dalitz plot in the
$K\rightarrow 3\pi$ decay. Thanks to the simultaneous collection of the two kaon charges and to the high resolution of the main sub-detectors, 
the systematics uncertainties are kept under the statistical error level. The experimental procedure, the analysis technique and the main systematics 
are discussed to present the final result 
$$
A_g=(1.8\pm2.6)\times 10^{-4}
$$
This result based on more than $45\cdot10^6$ events, correspondig to one half of the whole two year data taking,  is about an order of magnitude more precise with respect to the previous measurement.  
}
\section{Introduction}
In spite of the fact that more than 40 years have passed since the discovery of the CP violation~\cite{christ}, the full understanding of this
phenomenon is still far to be reached. After the discovery of direct CP violation in the neutral kaon decay into two pions in the
first part of the 90s, by the NA31 experiment~\cite{na31}, and the clear confirmation after few years, by 
NA48~\cite{na48} and KTEV~\cite{ktev}, the only other environment in which this effect was detected is the neutral B system~\cite{bdir}. 
On the other hand, the study 
of the tiny effects due to the violation of this symmetry in all the systems where it's possible, represents an important window on the 
contribution of new physics beyond the Standard Model (SM). 
In fact new effects could appear in a relevant way in the heavy quark loops which are at the core of the mechanism  
allowing CP violation in the decay. In the kaon sector the most promising places, besides $\varepsilon^{\prime}/\varepsilon$,
where this kind of contributions could play some role are the rates of GIM suppressed rare decays~\cite{D'Ambrosio:2001zh} and the charge asymmetry
between charged kaons.
In particular in $K\rightarrow 3\pi$ this asymmetry could give a strong qualitative indication of the validity of the CKM description 
of the direct CP violation or the existence of possible sources outside of this paradigm. Quantitative tests or predictions are, at the moment, very 
difficult from a theoretical point of view due to the complexity of the calculation involving non-perturbative hadronic effects.
 As soon as the intense theoretical
efforts to improve the understanding of this sector of the particle interactions will produce tools to make reliable predictions, the 
experimental measurements could also be used as strong quantitative test. 

From an experimental point of view, the easiest way to look for any possible differences between $K^+$ and $K^-$ decay into three pions is to compare
the \emph{shape} of the \emph{Dalitz plot} distribution instead of the decay rates. 
Exploiting the small Q value of the $K\rightarrow 3\pi$ decay 
it's possible to use a polynomial expansion of the matrix element \footnote{Due to the strong interaction process $\pi^+\pi^-\rightarrow\pi^0\pi^0$ this simple
parametrization isn't fully adequate to describe the matrix element~\cite{Batley:2005ax}. Nevertheless this expansion is adopted here for the CP violation measurement.}:
\begin{equation}\label{eqn:matrix}
|M(u,v)|^2\sim1+gu+hu^2+kv^2+... \quad \mbox{   ,}
\end{equation}
where g,h,k are called linear and quadratic slope parameters and where the Lorentz invariant Dalitz plot parameters~\cite{pdg} u and v
are related to the energy ($E_i^*$) of ``odd'' and ``even'' pion (i.e. the unpaired pion and the same sign two pions, respectively) in the 
center of mass, through the definition:
\begin{equation}
u=\frac{(s_3-s_0)}{m_{\pi}^2}=2m_K\frac{m_K/3-E^*_3}{m_{\pi}^2} \mbox{ , }
v=\frac{(s_2-s_1)}{m_{\pi}^2}=2m_K\frac{E^*_1-E^*_2}{m_{\pi}^2} \quad \mbox{   .}
\end{equation}
where $s_i=(p_K-p_i)^2$, being $p_K$ and $p_i$ the kaon and the pions (with $i=1,2,3$ and $i=3$ for the ``odd'' pion)  four-momenta, 
$s_0=(s_1+s_2+s_3)/3$, and $m_K$, $m_{\pi}$ the kaon and charged pion mass, respectively. The present value~\cite{pdg}
 for the linear slope \footnote{The linear slope for the v variable is forbidden for symmetry reasons.} appearing in (\ref{eqn:matrix}) are, for the two 
possible decay mode:
\begin{equation}
g(\pi^{\pm}\pi^0\pi^0)=0.638\pm0.020 \quad g(\pi^{\pm}\pi^+\pi^-)=-0.2154\pm0.0035 
\end{equation}
and ($|h|,|k|\ll |g|$). A difference between $g^+$ and $g^-$, the linear slopes in the $K^+$ and $K^-$ decay, represents a signal of direct CP violation.
This can be formalized defining the CP violating parameter $A_g$ as:
\begin{equation}\label{eqn:ag}
A_g=\frac{g^+-g^-}{g^++g^-}\sim\frac{\Delta g}{2g} \quad \mbox{   ,}
\end{equation}
where $\Delta g=g^+-g^-$ and $g$ is the average between $g^+$ and $g^-$. In the SM framework the prediction of the $A_g$ value is very difficult
due to hadronic effects contribution, but all the calculations~\cite{calcsm} are in the range 
between $10^{-6}$ and few $10^{-5}$, both in the \emph{neutral}
(the $K^{\pm}\rightarrow\pi^{\pm}\pi^0\pi^0$) and in the \emph{charged} (the $K^{\pm}\rightarrow\pi^{\pm}\pi^+\pi^-$) decay mode. Calculations
involving models beyond the SM allow for an enhancement up to the level of few $10^{-4}$. Several experiments in the past have searched for the
asymmetry in ``neutral'' and ``charged'' mode as summarized in tab.\ref{tab:exper}  . 
\begin{table}[!t]
\begin{center}
\begin{tabular}{|l|l|l|}
\hline
\emph{Asymmetry} & \emph{\# of events} & \emph{Experiment} \\
\hline
$A_g^0=(19\pm125)\cdot10^{-4}$ & 115K & CERN PS(1975) \cite{Smith:1973bi}\\
$A_g^0=(2\pm19)\cdot10^{-4}$ & 620K & Protvino IHEP (2005) \cite{Akopdzhanov:2004xb} \\
\hline
$A_g^c=(-70\pm53)\cdot10^{-4}$ & 3.2M & BNL AGS (1970)\cite{Ford:1970ta} \\
$A_g^c=(22\pm15\pm37)\cdot10^{-4}$ & 54M & HyperCP (2000) prelim.\cite{hypcp}\\
\hline
\end{tabular}
\end{center}
\caption{Summary of the experimental situation both in ``neutral'' ($A_g^0$) and in ``charged'' ($A_g^c$) mode, before the NA48/2 results}
\label{tab:exper}
\end{table}
The sensitivity reached so far is at level of few $10^{-3}$, dominated by the 
systematic uncertainties. The main goal of the NA48/2 experiment is to reach the sensitivity of $10^{-4}$ both in ``neutral'' and in ``charged'' mode, 
covering the gap existing between the experimental results and the ``beyond the SM'' theory predictions.

The NA48/2 experiment (basically an upgrade \cite{Batley:1999fv} of the NA48 experiment, devoted to the direct CP violation measurement
in the $K^0$ system) will be
briefly described, the analysis of the charged asymmetry in  $K^{\pm}\rightarrow\pi^{\pm}\pi^0\pi^0$ will be discussed and the final result, based 
on about a half of the whole NA48/2 data sample, will be presented.   
\section{NA48/2: beam line and experimental apparatus}
\begin{figure}[!t]\label{fig:beam}
\begin{center}
  \includegraphics[height=.2\textheight]{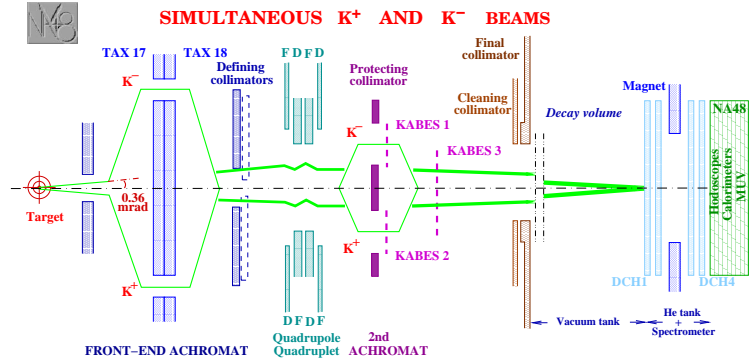}
\end{center}
  \caption{The NA48/2 beam line and detector. Not to scale.}
\end{figure}

A novel high intensity beam line was designed in order to allow for the simultaneous collections of $K^+$ and $K^-$ decays. 
The possibility to collect at the same time, with the same detector, the $K\rightarrow 3\pi$ decay of both charges, represents a fundamental 
point in the reduction of the systematic error. The beam transport line is schematically sketched in fig.\ref{fig:beam}.
The hadron beam, of both charges, is produced by the SPS 400 GeV/c protons ($\sim 7\cdot10^{11}$ protons per pulse) impinging on 
a 40 cm long and 2 mm diameter beryllium target with an angle of  zero degrees. A magnetic device, called \emph{first achromat}, selects
the momentum of the beam in the range $(60\pm3)$ GeV/c, splitting the two charges. After being recombined the beams are focused by a quadruplet of quadrupoles.
The \emph{second achromat} system houses a spectrometer for the beam, called KABES~\cite{peyo}, to measure the particles' momentum with a resolution of $1\%$, useful
in particular for rare decays studies. The beam, recombined again along the beam axis, contains $\sim6.4\cdot10^7$ particles per 4.8 s spill (about 12 times more pions than kaons)
with kaon charge ratio $K^+/K^-\sim1.8$, which is irrelevant for the measurement . In the 114 m long decay region the two beams, both $\sim$5 mm wide (RMS) are superimposed at level of 1 mm.
The central detector is based on the old NA48 detector described elsewhere~\cite{na48det}. The spectrometer magnet was operated in order to give 
$P_{\perp kick}=120MeV/c$ and the resolution in momentum (GeV/c) is $\sigma_p/p=1.0\%\oplus0.044\%\cdot p$. The LKr electromagnetic calorimeter is employed to
detect the photons from $\pi^0$ decay with an energy resolution (GeV) of  $\sigma_E/E=3.2\%/\sqrt{E}\oplus 9\%/E \oplus 0.42\%$. Thanks to these 
performances the resolution on the reconstructed kaon mass is good enough ($0.9 MeV/c^2$ in the ``neutral'' mode, $1.7 MeV/c^2$ in the ``charged'' mode) for a precise calibration of the detector instabilities. 

A two level trigger is employed to reduce the rate of data collected. A hardware level trigger (L1) using informations from the fast hodoscope counter
and from a dedicated LKr read out system is followed by a L2 section based on processors for fast reconstruction of kinematical quantities.
In particular for the $K^{\pm}\rightarrow\pi^{\pm}\pi^0\pi^0$ online selection, the missing mass of the charged pion is required to be far from the $\pi^0$
mass, in order to reduce the contribution of the more frequent $\pi^{\pm}\pi^0$ decay. 
The final trigger rate is 10 KHz.
\section{Strategy of the measurement}
The method to extract $A_g^0$ ($A_g$ for the ``neutral'' mode) is based on the comparison between the u projections of the Dalitz plot distribution. The
ratio between the two distributions is well described by:
\begin{equation}
R(u)=\frac{N_{K^+}}{N_{K^-}}\propto\frac{1+(g+\Delta g)u+hu^2}{1+gu+hu^2}\quad \mbox{   .}
\end{equation}
The presence of magnetic fields both in the beam sector (achromat) and in the detector (spectrometer magnet) introduces an intrinsic 
charge dependent acceptance of the apparatus. In order to equalize this asymmetry the main magnetic field are frequently reversed during the data taking 
(the spectrometer magnet on a daily basis, while the achromat system on a weekly basis). For each achromat polarity, it's possible to define two ratios
in which the same side (the two sides are called \emph{Jura} (J) and \emph{Saleve} (S)) of the spectrometer is involved:
\begin{equation}
R_J(u)=\frac{N_{K^+}^{B^-}}{N_{K^-}^{B^+}} \mbox{  ,  } \quad
R_S(u)=\frac{N_{K^+}^{B^+}}{N_{K^-}^{B^-}} \quad \mbox{   .}
\end{equation} 
In each of these ratios the acceptance asymmetry due to the spectrometer is cancelled. The full cancellation of the detector asymmetry is obtained
exploiting the quadruple ratio:
\begin{equation}\label{eqn:quad}
R(u)=R_{US}R_{UJ}R_{DS}R_{DJ}\sim \bar{R}(\frac{1+(g+\Delta g)u+hu^2}{1+gu+hu^2})^4 \quad \mbox{   ,}
\end{equation} 
where U and D stands for the path followed by the $K^+$ in the achromat (up and down)  and $\bar{R}$ is an inessential normalization constant. This method is independent from the relative size of the four samples collected 
with different fields configuration and from the $K^+$ and $K^-$ flux difference. The method allows for the cancellation of:
\begin{itemize}
\item local detector bias (left-right asymmetry), thanks to the fact that each single ratio in (\ref{eqn:quad}), is defined in the same side of the detector;
\item beam local biases, because in each single ratio the path of the particle through the achromat is the same;
\item global time variation biases, because the decays from both charges are collected at the same time.
\end{itemize}
The result remains sensitive only to the time variation of the detector with a characteristic time smaller than the inversion period of the magnetic field. In fact
each of the ratios employed in the (\ref{eqn:quad}) is constructed using data of subsequent days. Other systematic biases induced by effects not cancelled
by the alternation of the magnetic field (for instance the Earth's magnetic field and any misalignment of the spectrometer) have to be carefully corrected. 
The intrinsic cancellations of the acceptance asymmetries in the quadruple-ratio allow to avoid the use of simulation input for the measurement. Nevertheless a GEANT3~\cite{geant} based
MonteCarlo, including the time variation of beam and detectors, is used for systematic studies and as a cross-check of the result.
\section{Data Analysis in $K^{\pm}\rightarrow\pi^{\pm}\pi^0\pi^0$ mode}
The reconstruction of the event in the ``neutral'' mode exploits the LKr and the spectrometer.
In the preselection phase the tracks and the photons are selected according to quality criteria. The fiducial region of the detectors is chosen to avoid 
edge effects or miscalibrated regions. In particular the spectrometer's inner radial cut is chosen according to the actual beam position, whose displacement during the run is 
measured by three-track events. The decay vertex is recostructed from the $\gamma$ impact positions on the LKr, for each $\pi^0$, by using the relation:
\begin{equation}
Z_{ij}=\frac{1}{m_{\pi^0}}\sqrt{E_i E_j d_{ij}^2}\quad \mbox{   .}
\end{equation}
Among all the possible $\gamma$ pairings, the two pairs with minimum vertex difference are selected and the final decay vertex is obtained as the arithmetic average
of the two vertices. The kaon invariant mass is used to select the charged pion and to identify a good $K^{\pm}\rightarrow\pi^{\pm}\pi^0\pi^0$, by requiring  
$|M_{K^{\pm}}-M_{K^{\pm}_{PDG}}|<6 MeV/c^2$. 
The measurement of the charged pion momentum, slightly biased by variable DCH misalignment, is corrected by exploiting the condition $M_{K^+}=M_{K^-}$
in the reconstruction of the $K^{\pm}\rightarrow\pi^{\pm}\pi^+\pi^-$ mode. The same decay mode is used to follow the variation of the absolute value of the
spectrometer magnetic field after each periodical inversion, monitored online only at level of $\sim10^{-3}$. The position of the photons impact point
is corrected to take into account the projective geometry of the calorimeter. These effects gives, after the correction, a negligible contribution to the systematics
error. 

A potential source 
of systematic bias is the trigger. This was studied directly from the data with downscaled sample of control triggers, uncorrelated with the main trigger.
In the ``neutral'' mode the L1 trigger level is essentially composed by a coincidence of a signal in the scintillating hodoscope (Q1) and a pattern
 compatible with 4 photons on the LKr (NTPEAK). The Q1 efficiency is obtained by using all the events with one track in the charged hodoscope,
re-weighted according to the distribution of the $3\pi$ events. The inefficiency ($\sim0.25\%$) is constant during the run  and the upper limit of the
systematic error on $\Delta g$ given by this part of the trigger is $0.1\cdot10^{-4}$. 

The largest systematic uncertainty ($1.3\cdot10^{-4}$) comes from the neutral part
of the L1. The NTPEAK inefficiency is different at the beginning ($\sim0.7\%$) and at the end ($\sim3\%$) of the data taking. The estimation
of this systematics is limited by the statistics of the control sample. The L2 inefficiency ($\sim5.7\%$ on average during the run) is mostly  ($\sim70\%$) due to the DCH wires inefficiencies and was studied by using MonteCarlo simulation, including the detailed time variation of the spectrometer's behaviour. The
systematic error assigned to $\Delta g$ is $0.4\cdot10^{-4}$, including also other effects like data buffer overflows, algorithm inefficiency and synchronization effects in the
trigger chain. All the sources of systematics are summarized in table \ref{tab:syst} .
\begin{table}[!t]
\begin{center}
\begin{tabular}{|l|r|}
\hline
\emph{Systematic effect} & \emph{Effect on $\Delta g\cdot 10^{-4}$} \\
\hline
\hline
U calculation \& fitting & $\pm0.4$ \\
LKr non linearity& $\pm0.1$ \\
Shower overlapping & $\pm0.5$ \\
Pion decay & $\pm0.5$ \\
Spectrometer Alignment \& Momentum scale & $<\pm0.1$ \\
Beam Geometry & $\pm0.3$ \\
Accidentals & $\pm0.2$ \\
L1 Trigger: Q1 & $\pm0.1$ \\
L1 Trigger: NTPEAK & $\pm1.3$(stat.) \\
L2 Trigger & $\pm0.4$ \\
\hline
Total systematic uncertainty & $\pm1.0$ \\
Trigger statistical uncertainty & $\pm1.3$ \\
\hline
\end{tabular}
\end{center}
\label{tab:syst}
\end{table}

During the 2003 data taking (one half of the whole statistics)  $40.3\cdot10^6$ $K^+$ and $16.9\cdot10^6$ $K^-$ decays in the ``neutral'' $3\pi$ mode have been selected for the
asymmetry measurement presented in this paper. The final result in terms of $A_g$ (related to $\Delta g$ through (\ref{eqn:ag}) ) , 
obtained as average of three independent analyses that agree within few $10^{-5}$, is:
\begin{equation}\label{eqn:resul}
A_g^0=(1.8\pm2.2_ {stat.}\pm1.0_{stat.(trig)}\pm0.8_{syst.}\pm0.2_{ext.})\cdot10^{-4}=(1.8\pm2.6)\cdot10^{-4} \quad \mbox{   ,}
\end{equation}  
where the external error depends on the uncertainties on the value of $g(\pi^{\pm}\pi^0\pi^0)$ ($\sim 3.1\%$)~\cite{pdg}.
This result, fully compatible with the SM predictions, is almost one order of magnitude better than the previous measurement by other experiments~\cite{Smith:1973bi},~\cite{Akopdzhanov:2004xb}.
The ongoing analysis of the 2004 sample will improve the statistical error of a factor of $\sim1.4$, with a similar reduction in the systematic error.

\section{Result in charged mode} 
An analogous analysis in the ``charged'' mode using $1.6\cdot10^9$ $K^{\pm}\rightarrow\pi^{\pm}\pi^+\pi^-$ decays, from the 2003 sample, gives the result~\cite{Batley:2006mu}:
\begin{equation}
A_g^c=(1.7\pm2.1_{stat}\pm1.4_{trig}\pm1.4_{syst})\cdot10^{-4}=(1.7\pm2.9)\cdot10^{-4}\quad \mbox{   .}
\end{equation}
The statistical precision is similar with respect to the result given in  (\ref{eqn:resul}) because, even though the ``neutral'' mode is statistically disfavored 
due to lower branching ratio and acceptance, the population density of the Dalitz plot is more favourable and $|g(\pi^{\pm}\pi^0\pi^0)|\sim3\cdot|g(\pi^{\pm}\pi^+\pi^-)|$.
A preliminary study on the combined 2003 and 2004 data ($3.11\cdot10^{11}$ events) leads to a compatible result~\cite{evgueni}:
\begin{equation}
A_g^c=(-1.3\pm1.5_{stat}\pm0.9_{trig}\pm1.4_{syst})\cdot10^{-4}=(-1.3\pm2.3)\cdot10^{-4}\quad \mbox{   .}
\end{equation}

\end{document}